\documentclass[aps,pre,twocolumn,showpacs]{revtex4}
\usepackage{epsfig}
\usepackage{times}
\begin{document}

\title{Three-state Potts model in combination with the rock-scissors-paper game}

\author{Attila Szolnoki, Gy\"orgy Szab\'o,}
\affiliation
{Research Institute for Technical Physics and Materials Science
P.O. Box 49, H-1525 Budapest, Hungary}
\author{M\'aria Ravasz}
\affiliation
{Babe\c{s}-Bolyai University, Department of Physics, RO-400084 Cluj, Romania}

\begin{abstract}
We study a three-state Potts model extended by allowing cyclic dominance 
between the states as it appears for the rock-scissors-paper game.
Monte Carlo simulations are performed on a square lattice when varying
the temperature and the strength of cyclic dominance. It is shown that
the critical phase transition from the disordered state to the ordered
one is destroyed by the cyclic dominance that yields a self-organizing 
pattern even at low temperatures. The differences and similarities are 
discussed between the present model and the half-filled, driven lattice
gases with repulsive interaction.
\end{abstract}


\maketitle

The ordering phenomena and the related phase transitions are already
well understood in the equilibrium systems \cite{stanley:71} while
the theoretical understanding of the non-equilibrium phase transitions
is still at its beginning \cite{marro:99,hinrichsen:ap00}. 
Many relevant and general features of these transitions in the equilibrium
systems can be studied by the Potts models \cite{wu:rmp82}. Now we
introduce an extended version of the three-state Potts model to
investigate the effect of cyclic dominance between the states. This
model can be considered as a combination of the traditional Potts
model and a spatial rock-scissors-paper game (sometimes called as
three-state cyclic predator-prey or Lotka-Volterra models). The strength
of cyclic dominance will be characterized by a single parameter ($\varepsilon$)
in such a way that for $\varepsilon = 0$ the system becomes equivalent to
the equilibrium Potts model exhibiting a well-known critical transition.
The consideration of this model was strongly
motivated by the work of Katz et al. \cite{katz:prb83,katz:jsp84} who
introduced the concept of driven lattice gases to study the effect of
an external electric field on the ordering process. Since the appearance
of their pioneering works many general features of these systems have
already been explored (for a review see the 
Refs. \cite{schmittmann:95,marro:99}). 

The present model provides a continuous transition between an equilibrium
system and a spatial evolutionary game where the time-reversal symmetry
(detailed balance) is broken at the elementary (microscopic) steps.
It will be shown that in the presence of cyclic dominance the long-range
order cannot be observed even for low temperatures, and thereby the 
critical transition is also suppressed. A similar phenomenon has already
been observed for the two-dimensional driven lattice gas with repulsive
interactions where the formation of long-range order is prevented by
an interfacial instability due to the enhanced particle transport
along the boundaries separating the "chessboard" and "anti-chessboard"
ordered phases \cite{szabo:pre94,szabo:pre97}. 

Our analysis is focused on a two-dimensional system where each site $x$
of a square lattice is characterized by a three-state site variable,
namely, $s(x)=s_0$, $s_1$, and $s_2$. For later convenience these states
(strategies, species, etc.) will be denoted by the basis vectors of a
three-dimensional space, i.e.
\begin{equation}
s_0=\left( \matrix{1 \cr 0 \cr 0} \right) \;,\;\;\; 
s_1=\left( \matrix{0 \cr 1 \cr 0} \right) \;,\;\;\; 
s_2=\left( \matrix{0 \cr 0 \cr 1} \right) \;.
\label{eq:states}
\end{equation}
The time evolution of the system is governed by random sequential updates. 
More precisely, the transition probability from a state $s(x)$ (at site
$x$) to a randomly chosen state $s^{\prime}(x)$ is given as
\begin{equation}
W[s(x) \rightarrow s^{\prime}(x)] = {1 \over 1 + \exp{(-\delta U(x) / T)}}
\label{eq:w}
\end{equation}
where $\delta U(x)$ is the difference of payoffs between the final and 
initial states, and $T$ is the temperature characterising the effect of 
the noise. The payoff at site $x$ depends on $s(x)$ as well as on the
neighboring states [$s(y)$] as given by the following sum of matrix 
products:
\begin{equation}
U(x) = \sum_{<y>} s^{+}(x) A s(y)
\label{eq:u}
\end{equation}
where the summation runs over the nearest-neighbors of the site $x$,
$s^{+}(x)$ denotes the transpose of $s(x)$, and the payoff matrix
$A$ is defined as
\begin{equation}
A = \left(\matrix{ 1        & \varepsilon  & -\varepsilon \cr
                  -\varepsilon &  1        &  \varepsilon \cr
                   \varepsilon & -\varepsilon &  1       }\right) \;.
\label{eq:A}
\end{equation}
In the limit $\varepsilon \to 0$ this model can be considered as a (kinetic)
three-state ferromagnetic Potts model \cite{wu:rmp82} with a Glauber
dynamics \cite{glauber:jmp63}. Evidently, in this case the total energy
is defined as $H=-\sum_x U(x)/2$ and the microscopic processes satisfy the
detailed balance in equilibrium. Consequently, the system tends towards a
stationary state whose statistical features are described by the Gibbs
ensemble. When decreasing the temperature the Potts model undergoes an
ordering process from the disordered state to one of the three equivalent
homogeneous (ordered) states. The corresponding critical transition
represents a well-known universality class \cite{wu:rmp82,stanley:71}.

For $\varepsilon > 0$ the off-diagonal components of the payoff matrix
$A$ are asymmetrical therefore the total payoff (or the above defined 
$H$) is not affected by the value of $\varepsilon$ for any states.
At the same time, the value of $\varepsilon$ influences the probability
of strategy changes because $W[s(x) \rightarrow s^{\prime}(x)]$ depends
definitely on the variation of individual payoff [$ \delta U(x)$] and the
above evolutionary rule manifests a way how the (selfish) individuals
wish to maximize their own payoff without any concern about the neighbors'
performance. As a result, cyclic invasions occur
along the boundaries separating homogeneous domains; domains of state
$s_0$ are invaded by $s_1$ invaded by $s_2$ invaded by $s_0$. These
cyclic invasions are capable to maintain a self-organizing pattern
with rotating spiral arms whose ``velocity'' is controlled by
$\varepsilon$ \cite{tainaka:pre94,frachebourg:pre96a,szabo:pre99}.

For most of the spatial evolutionary games the choice of the dynamical
rules (or $W[s(x) \rightarrow s^{\prime}(x)]$) is based on a learning 
mechanism or strategy adoption modelling the Darwinian selections
\cite{nowak:jbifchaos93,hauert:jbifchaos02,szabo:pre98}. In these
models different ways are suggested for the players to adopt the
strategy of their more successful neighbors. The common feature of
these strategy adoptions is that the new state will be equivalent to
one of the neighboring strategies. Consequently, this mechanism prohibits
the variation inside the homogeneous domains and makes the extinction
process to be similar to those defined by the contact process
(or directed percolation) \cite{marro:99,hinrichsen:ap00}.
In the present model, however, the above ``Glauber dynamics'' allows 
the players to choose all the possible strategies therefore the time
variation is not restricted to the interfaces separating the homogeneous
domains. In the context of evolutionary game theory the above
evolutionary rule describes a different behavior. Namely, here the
players know all their possibilities and their choices depend on the
increase of income what they are able to evaluate in the knowledge
of neighboring strategies.

In the present work we study the effect of the "cyclic dominance" 
on the phase transition. For this purpose systematic Monte Carlo (MC)
simulations are performed on a square lattice under periodic boundary
conditions varying the temperature $T$ and strength $\varepsilon$ of
the cyclic dominance for different linear sizes $L$. Each simulation
is started from a random initial state and after a suitable thermalization
time we have recorded the concentration of states ($\rho_0$, $\rho_1$,
and $\rho_2$) for each Monte Carlo steps. 
We have also made simulations starting from ordered homogeneous phases
to check the stability of the stationary state.
To investigate the ordering process, we have determined the average value 
of the order parameter from the values of concentration data 
\cite{wu:rmp82,challa:prb86} 
\begin{equation}
m = {1 \over 2} \langle [3 \max(\rho_0,\rho_1,\rho_2) -1 ] \rangle
\label{eq:op}
\end{equation}
where $\langle \cdots \rangle$ refers to averaging over a sampling
time varied from $10^5$ to $10^6$ Monte Carlo steps per sites (MCS).
In the disordered phases $\rho_0=\rho_1=\rho_2 = 1/3$ (due to the
cyclic symmetry) and $m=0$ in the thermodynamic limit ($L \to \infty$).
For $\varepsilon = 0$ and below the critical temperature ($T<T_c=0.995(1)$)
the system evolves into one of the long-range ordered (symmetry breaking)
stationary states (e.g., $\langle \rho_0 \rangle =(1+2m)/3$ and 
$\langle \rho_1 \rangle = \langle \rho_2 \rangle =(1-m)/3$ and the
remaining two equivalent states are given by the cyclic permutation of
indices) if the linear size is sufficiently large. In the thermodynamic
limit the order parameter $m$ decreases monotonously from 1 to 0 as
the temperature is increased from 0 to $T_c$ and the vanishing of $m$
follows a power law behavior if $T_c$ is approached from below 
\cite{wu:rmp82}. For finite sizes, however, the MC simulations 
exhibit a smoothed order parameter function that deviates monotonously
if we decrease the system size \cite{challa:prb86,landau:00}. 
Significantly different
finite size effects are observed when investigating the present model
for $\varepsilon >0$.

\begin{figure}[h]
\centerline{\epsfig{file=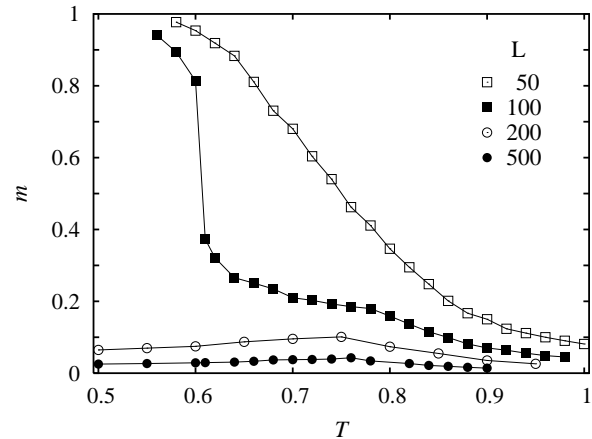,width=8cm}}
\caption{\label{fig:jop}Monte Carlo data for the order parameter {\it vs}. 
temperature at $\varepsilon =0.1$ for different system sizes as indicated
The solid lines are guides to the eye.}
\end{figure}

\begin{figure}[h]
\centerline{\epsfig{file=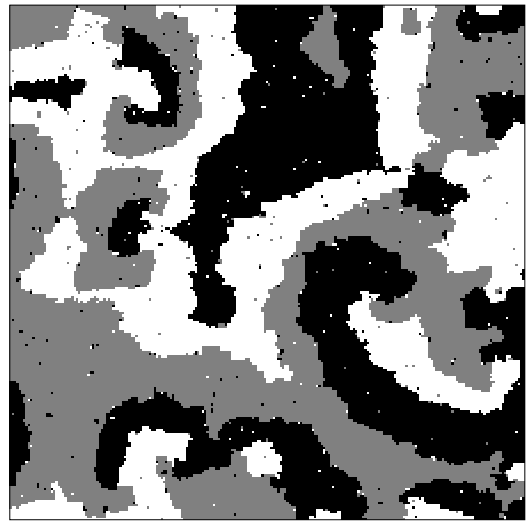,width=7cm}}
\caption{\label{fig:rspdist}Snapshot on a typical domain structure appearing
for $T=0.64$ and $\varepsilon = 0.1$. The three-edge vortices (antivortices) 
rotate in clockwise (anticlockwise) direction with spiral arms because
the average velocity of the invasion fronts (white invades black invades
gray invades white) is hardly affected by their curvature.}
\end{figure}

Figure ~\ref{fig:jop} illustrates how the order parameter $m$ varies 
with temperature $T$ for different linear sizes if $\varepsilon = 0.1$.
Apparently the MC data refers to an ordering process for small sizes 
($L=50$ and 100) bearing a resemblance to MC data obtained for
$\varepsilon = 0$. On the contrary, for example when $L=500$, the MC data 
do not indicate the appearance of long-range order. Instead of it a
self-organizing, three-color domain structure can be observed when
visualizing the time-dependence of the spatial distribution (for
a snapshot, see Fig.~\ref{fig:rspdist}).

On this snapshot one can identify all the three ordered phases forming
domains with a characteristic linear size $l$. For $T<T_c$ and $\varepsilon=0$
the growth of these domains ($l \sim \sqrt{t}$) is driven by the interfacial
(Potts) energy \cite{grest:prb88,mouritsen:impb90,bray:ap94}. Here, however, this domain growth is
prevented by two processes emerging for $\varepsilon > 0$. The first process
is related to the appearance of rotating spiral arms for the three-edge
vortices where the three types of domains (or domain boundaries) meet.
On these maps we can distinguish vortices and antivortices rotating
in opposite directions. Some topological and geometrical features of
such spatio-temporal patterns were already investigated in previous
papers \cite{tainaka:pre94,frachebourg:pre96a,szabo:pre99,szabo:pre02a}.
It is found that the spirals become well-marked if a ``surface tension''
is switched on and then the corresponding patterns cannot be characterized
by a single parameter (e.g., typical domain size or correlation
length) \cite{szabo:pre02a}.

In the present model there exists a second process causing the appearance
of growing domains via a nucleation mechanism inside the large
``homogeneous'' territories. Due to this process an ``ordered state'' 
prevailed by $s_0$ will be transformed into another one prevailed by 
$s_1$ as indicated by simulations for small sizes. In these cases the
three ordered states follow each other cyclically and the above method 
yields a sufficiently large value for $m$ (see Fig.~\ref{fig:jop}). 
Both the duration time and the probability of these transitions, that 
are initiated by a nucleation mechanism due to the thermal fluctuations,
increase with the system size. This is the reason why the values of $m$
are higher for $L=50$ than those for $L=100$ in the Fig.~\ref{fig:jop}.

For sufficiently large system sizes ($L >> l$) both mentioned mechanisms
work simultaneously and result in a self-organizing pattern where the
three states are present with the same concentration and {$m=0$). Henceforth
the quantitative investigations will be focused on the large systems ($L>500$)
and to the region of temperature ($T>0.6 T_c$) where the spatial 
patterns are isotropic.

Now we study the variation of correlation length $\xi$ derived from the 
asymptotic behavior (exponential vanishing) of the two-site correlation
function \cite{szabo:pre94}. For this purpose a series of MC simulations is
performed by varying the temperature for $\varepsilon=0.1$. The inset
in Fig.~\ref{fig:ksi} illustrates the absence of divergence in $\xi$ as
expected. When decreasing the temperature the correlation length
increases monotonously until a maximum value ($\xi \simeq 11.5(5)$
measured in lattice unit). Below the peak at $T \simeq 0.77$ the 
visualization of the distribution of species shows a self-organizing 
pattern (see Fig.~\ref{fig:rspdist}) and here the value of $\xi$ 
decreases very slowly with $T$. 

\begin{figure}[h]
\centerline{\epsfig{file=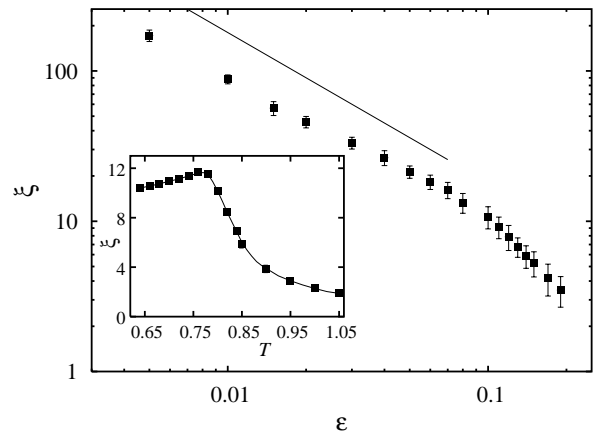,width=8cm}}
\caption{\label{fig:ksi}Variation of correlation length (with error bars)
as a function of $\varepsilon$ for fixed temperature ($T=0.75$) The straight
line indicates a power law divergence with a slope of $-1$. 
Inset shows the correlation length {\it vs.} temperature for 
$\varepsilon = 0.1$.}
\end{figure}

The $\varepsilon$-dependence of the correlation length is also determined
for a fixed temperature and the results are illustrated in a log-log
plot (see Fig.~\ref{fig:ksi}). These MC data are consistent with a
prediction $\xi \sim 1/\varepsilon$ for small $\varepsilon$. In the
typical size of domains similar divergence was found previously for a 
model where the nucleation mechanism was blocked \cite{szabo:pre02a}. 
This observation refers to a minor role of nucleation mechanism in the
maintenance of the self-organizing patterns at 
sufficiently low temperature
($T < T_c$). For higher temperature the nucleation mechanism plays a
crucial role by preventing the formation of monodomain state even
for $\varepsilon = 0$. Thus it is conjectured that this process results
in a different behavior of $\xi$ at the critical temperature.
Furthermore, here it is worth mentioning that in the above mentioned 
driven lattice gas model the transversal correlation length was also
proportional to the inverse of the strength of driving field 
\cite{szabo:pre97}.

In such systems the Potts energy measures the concentration
of domain walls and it gives an additional information about spatial
distributions. From the average Potts energy as a function of
temperature one can derive a specific heat ($c=d \langle H \rangle / dT$)
that exhibits a $\lambda$-divergence at the critical temperature in the
equilibrium limit ($\varepsilon=0$). Figure~\ref{fig:c} illustrates how
the $\lambda$-divergence is smoothed out if the cyclic dominance is
switched on. When choosing larger and larger $\varepsilon$ the maximum
value of specific heat decreases meanwhile the peak position moves towards
the lower and lower temperatures. 
The peaks are so shallow in the "driven" cases that a logarithmic scale
was necessary to present them in the same figure.
The appearance of the this peak
in the specific heat can be interpreted as a sign of the short range
ordering.

\begin{figure}[h]
\centerline{\epsfig{file=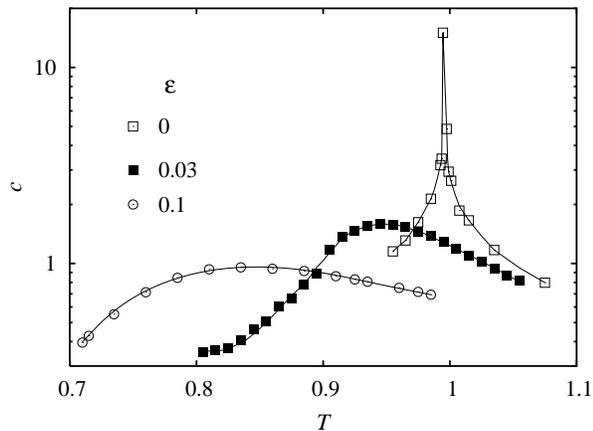,width=8cm}}
\caption{\label{fig:c}Specific heat as a function of temperature for 
three different values of $\varepsilon$ as indicated.
These MC data are obtained for such a large linear size ($L=1000$) 
where the size effects are already negligible.}
\end{figure}

A similar phenomenon was observed for the driven
lattice gases with repulsive interaction when increasing the external 
electric field \cite{szabo:pre94,szabo:pre97}. Although the observed
patterns and microscopic mechanisms are very different, in both cases 
interfacial effects prevent the formation of long-range order in the
presence of driving force. In these cases the interfaces belong to
the stationary states and their geometrical characterization requires
additional parameters. This general feature can occur in some other
non-equilibrium systems (e.g., in ecological models) where an external 
force induces some extra activity along the interfaces separating the
"ordered domains". 

In summary, a three-state dynamical lattice model is introduced by
combining the Potts model and the rock-scissors-paper game
to study the effect of cyclic dominance on the ordering process.
Due to the cyclic dominance the time-reversal symmetry is broken at the
elementary steps and thereby the behavior of this this model cannot be
described by the
methods of equilibrium statistical physics. Our numerical analyses have
justified that both the long-range (symmetry breaking) ordering process
and the corresponding critical transition are suppressed in the presence
of cyclic dominance ($\varepsilon > 0$). According to the simulations
the three equivalent ordered phases coexist by forming a self-organizing 
domain structure even at low temperatures and sufficiently weak cyclic
dominance. The equilibrium state is approached via the divergence of the
typical domain size when the strength of cyclic dominance goes to zero.

\begin{acknowledgments}

This work was supported by the Hungarian National Research Fund under
Grant No. T-47003.

\end{acknowledgments}

\end{document}